# Terahertz nano antenna enabled early transition in VO$_2$

Young-Gyun Jeong[1], Jae-Wook Choi[1], Sang-Hoon Han[2], Ji-Soo Kyoung[1], Hyeong-Ryeol Park[1], Namkyoo Park[2], Bong-Jun Kim[3], Hyun-Tak Kim[3] and Dai-Sik Kim[1*]

[1]Department of Physics and Astronomy and Center for Subwavelength Optics, Seoul National University, Seoul, 151-747, Republic of Korea

[2]Photonic Systems Laboratory, School of EECS, Seoul National University, Seoul 151-744, Republic of Korea

[3]Metal-Insulator Transition Creative Research Center, Electronics and Telecommunications Research Institute, Daejeon, 305-700, Republic of Korea

[*]dsk@phya.snu.ac.kr

We study terahertz transmission through nano-patterned vanadium dioxide thin film. It is found that the patterning allows the lowering of the apparent transition temperature. For the case of the smallest width nano antennas, the transition temperature is lower by as many as ten degrees relative to the bare film, so that the nano patterned hysteresis curves completely separate themselves from their bare film counterparts. This early transition comes from the one order of magnitude enhanced effective dielectric constants by nano antennas. This phenomenon opens up the possibility of transition temperature engineering.

Insulator-to-metal phase transition materials have been an interesting research subject with enormous application potentials due to their drastic changes in optical, electronic, structural properties [1]. As one good example, vanadium dioxide ($VO_2$) displays first-order insulator-to-metal phase transition near 68 $^oC$ [2]. This transition can also be triggered by optical beam, electric bias or external strain [3-6]. The lattice structure is transformed from monoclinic to tetragonal structure as the temperature rises. Also, the dielectric constant changes by several orders of magnitude during the transition [7].

In near infrared (IR)-terahertz (THz) regime, $VO_2$ shows dynamic switching behavior through the insulator-to-metal transition [8, 9]. To develop an active switching device with $VO_2$, improved switching mechanism and efficiency are required. Previously, active full control of THz transmission through $VO_2$ thin film combined with slot antenna array type gold nanostructure have been reported with greatly enhanced switching efficiency [10-14].

For the more practical applications, there have been many approaches to reduce the transition temperature of $VO_2$ such as doping, different substrate type and various fabrication techniques [15-21]. In this Letter, we found a new approach towards lowering the apparent transition temperature in $VO_2$ without changing the material parameters themselves. We demonstrate the thermal hysteresis curves shifting towards lower temperature, measured by THz time domain spectroscopy, when THz nano-width slot antennas are fabricated on $VO_2$ thin film. Introduction of pattern-induced large-k wave vectors, which respond much more sensitively to the changing dielectric constants of the material, is responsible.

Our 100-nm-thick $VO_2$ films were grown by the pulse laser deposition (PLD) method on 430 μm thick C-plane sapphire substrate [22]. Electron beam lithography with negative photoresist and single-layer lift-off process is applied to fabricate nano antenna array pattern on $VO_2$ [10]. The widths of nano antennas are varied between 120 nm and 2.5 μm, with other parameters fixed. Figure 1 (a) represents our sample schematic and SEM image of typical nano antennas. The total area of the nano antenna array is 2 cm by 2 cm and the length of each antenna is 150 μm. The adjacent antennas are separated by 10 μm in the length direction and 30 μm in the width direction. Figure 1 (b) shows THz transmission spectra of 380 nm-width nano antenna array patterned 100 nm-thick $VO_2$ film. It is normalized by the bare film transmission signal. Field enhancement due to energy funneling of nano antenna array is calculated from the nano antenna coverage ratio which is around 1 % [23, 24]. At the resonance frequency, the enhancement is more than 60 due to shape resonance with the large aspect ratio of around 500.

We measure THz transmission signal through bare and nano antenna array patterned $VO_2$ films at different temperatures. First we heat the sample from 30 °C to 100 °C and cool it down back. To compare hysteresis curves amongst them, we plot temperature dependant normalized THz transmission amplitude at 0.5 THz. Our interests are focused on the transition temperature so that we normalize the maximum transmission amplitude as 1 and set the minimum at 0 using an offset. In heating process, the THz transmission signal starts to decrease at around 68 °C in bare film due to the insulator-to-metal phase transition of $VO_2$. However, when 120 nm-width nano antenna array is patterned on the bare film, the THz transmission starts to decrease from 30 °C and the hysteresis curve shows much lower transition temperature, defined as the mid-point of the transition curve. One possible cause of early THz transmission change in temperature domain is the incident THz field amplitude which may induce nonlinear effect due to large field enhancement. In the inset of Figure 1 (c), we measure the transition temperature from thermal hysteresis curves of PLD type bare and 380-nm-width nano antenna patterned $VO_2$ film according to the incident THz field amplitude. We observe that the transition temperature shift on nano antenna array patterned $VO_2$ film does not depend on the incident THz field amplitude, as expected owing to the small intensity of the THz beam, indicating that this is a linear, purely pattern-induced effect.

In Figure 1 (d), we display the width-dependence of the transition temperature, defined as the half-point of the hysteresis curve, both for the heating process (red triangle), and the cooling process (blue inverted triangle); the black squares represent the average. It is clear that the narrower the line width, the lower the transition temperature, opening up the exciting possibility of transition temperature engineering.

To verify the generality of the early transition with nano-patterning, we examine other $VO_2$ films grown by different fabrication method, film thickness and on different substrate. In Figure 2 (a), the 120-nm-thick bare film is fabricated by RF-magnetron sputtering technique [25]. The early transition phenomenon remains intact despite having a different growth mechanism. When the bare film thickness becomes twice, still we observe the complete separation of the patterned and un-patterned THz transmission hysteresis curves (Fig. 2 (b)), showing again the generality of pattern-induced lowering of the transition temperature. In the case of much thinner 20 nm film grown by the sol-gel method [26], though the 200-nm-width nano antenna is narrower than other nano antennas in thicker films, the amount of shifted transition temperature is smaller so that the hysteresis curves overlap somewhat (Fig. 2 (c)). This means that the nano antennas are less functional in the ultra-thin film case.

It has been reported that insulator-to-metal transition of $VO_2$ has substrate orientation dependence [19, 27, 28]. In sapphire substrate, there are four different types of orientation, C-plane, A-plane, M-plane and R-plane. We measure THz transmission hysteresis curve through 120-nm-thick $VO_2$ film fabricated on the R-plane sapphire substrate (Fig. 2 (d)). The results show overall the same trends, except that there exists about four degrees shift towards lower temperature for both patterned and un-patterned films relative to the C-plane sapphire substrate. Having established the generality of the early transition regarding growth methods, substrate and film thickness, we now perform theoretical calculations to understand the essential physics.

As a theoretical approach, we compute temperature dependent normalized THz transmission spectra with the finite-difference time-domain (FDTD) method (Fig. 3 (a)), assuming bulk dielectric constants from earlier works [7, 29-34]. In FDTD calculation, the antenna widths vary from 2 μm to 10 nm. The FDTD results clearly reproduces that the transition happens at lower temperature than the bare film in nano antenna array pattered $VO_2$ sample, the narrower the width, the lower the transition temperature, despite maintaining the same bulk $VO_2$ parameter. Though we observe ten degrees shift from 120-nm-width antenna array patterned $VO_2$ sample in experimental results, the FDTD results suggest that more than fifteen degrees shift is possible if we could pattern sub 10-nm-width nano antennas. From the FDTD result of THz transmission, we calculate the effective dielectric constants of the nano antenna array-$VO_2$ composite/metamaterial as a function of temperature (Fig. 3 (b)) [35, 36]. The dielectric constant of the bare $VO_2$ film changes by four orders of magnitude in THz regime as we go from the insulator to metallic; however, the effective dielectric constant of our nano antenna patterned $VO_2$ film is increased another order of magnitude, up to five orders of magnitude change in sub-10-nm-width case. This suggests a remarkable potential to engineer the effective dielectric constants of the metal nanostructure-$VO_2$ composite, possibly up to those of the metal itself. Figure 3 (c) shows the width-dependent transition temperature obtained from the FDTD calculation (black circles), in good agreements with experimental data (red circles); the dashed line is the guide to the eye. The calculations show that the near-field absorption enhancement, orders of magnitudes larger than the bulk case of the same dielectric constants, is one important factor contributing to the earlier transition.

To further gain insights into the physics, in Figure 3 (d), first we plot the temperature dependent THz transmission curve of bare $VO_2$ from using well-known analytic, multiple-interference expressions for a plane parallel absorbing film situated between two dielectric media model [37]. Next, we model the case of the nano antenna emphasizing the introduction of large k-wave vectors that inevitably accompanies extreme angle diffraction from the aperture. Of these wave vectors, many are lost from the view point of transmission, owing to the increasing fraction of total internal reflection with increasing index of refraction by heating. In bare $VO_2$, because the incident THz wave maintains normal direction during it passes through the interfaces, total internal reflection loss does not factor in the multiple interference process, even as the index of refraction increases, limiting the sensitivity of bare film transmission to radically changing index of refraction. In addition in bare films, the reflected beams interfere constructively each other so that the THz transmission does not decrease as much with increasing index because the film thickness is much smaller than the wavelength. Therefore the early transition phenomenon can be explained in the following way: first, when the $VO_2$ film is insulating, energy funneling occurs through the nano apertures so that the patterned sample is more or less transparent [23]. As the index of refraction increases, the critical angle loss becomes very important as well as the absorption loss, which makes the transmission to respond much more sensitively to the changing dielectric constant compared to bulk, where the multiple interference limits the sensitivity.

From our experimental and theoretical results, we can modulate phase transition temperature of $VO_2$ by more than 10 degrees without changing the original material properties. Moreover, when we consider the generality of this early transition phenomenon, the transition temperature can be further lowered in combination with other techniques such as tungsten doping, external strain, and substrate treatment. Not only $VO_2$ but also other phase transition materials are good candidates for the transition temperature engineering in terahertz regime.

In conclusion, we demonstrate that THz waves transmitting through nano antenna array patterned $VO_2$ thin film is switched off at substantially lower temperature than 68 $^o$C which is the original insulator-to-metal transition temperature of bare $VO_2$ film. The transition temperature depends on the width of nano antennas and the shifted amounts are inversely proportional to the width. This can be interpreted as the amplification of the effective dielectric constants, up to one order of magnitude, by decreasing the width of nano antennas. This has wide potential to develop dielectric constant modulated near room temperature phase transition devices combined with nano structures.

This work was supported by the National Research Foundation of Korea (NRF) grant funded by the Korean government (MEST) (SRC, No. 2008-0062254) (Nos. 2010-0029648, 2010-0028713, 2011-0019170, 2011-0020209), KICOS (GRL, K20815000003), the creative research project of ETRI.

**Figure 1.** (a) Sample schematic: Each antenna is 150 μm in length and the width is varied (min. 120 nm to max. 2.5 μm). The antennas are separated by 10 μm in the vertical direction and the period is 30 μm in the horizontal. (b) THz transmission spectra and field enhancement factor of 380 nm-width nano antenna array patterned 100 nm-thick $VO_2$ film.. (c) Thermal hysteresis curve of normalized THz transmission amplitude of pulsed laser deposition (PLD) type 100-nm-thick bare and 120-nm-width nano antenna array patterned $VO_2$ film at 0.5 THz. The triangle indicates the heating process and the inverted triangle means the cooling. (inset) Incident THz field intensity dependant transition temperature of bare and 380-nm-width nano antenna array patterned $VO_2$ film at 0.5 THz. (d) Nano antenna width dependant transition temperature, defined as the average of the cooling and heating transitions.

**Figure 2.** Thermal hysteresis curves of normalized THz transmission amplitudes measured at 0.5 THz for bare (black triangles) and patterned (red triangles) $VO_2$ samples of various growth methods: (a) RF-magnetron sputtering type 120-nm-thick film with 300-nm-width nano antennas. (b) a sputtering type 240-nm-thick film with 320-nm-width nano antennas. (c) sol-gel type 20-nm-thick film with 200-nm-width nano antennas. (d) sputtering type 120-nm-thick film with 350-nm-width nano antennas, now on an R-plane sapphire substrate.

**Figure 3.** (a) FDTD simulation results of bare and various nano antenna arrays with the same parameters with experiments. The 'w' means the width of nano antennas in each case. (b) Effective dielectric constants obtained from the FDTD result. (c) Transition temperature deduced FDTD simulations (black circles) together with experimental results (red circles). (d) Analytic model calculation of bare and an ultra-narrow nano antenna cases. The funneled THz wave diffracts to a wide angle on the substrate side.

Fig. 1

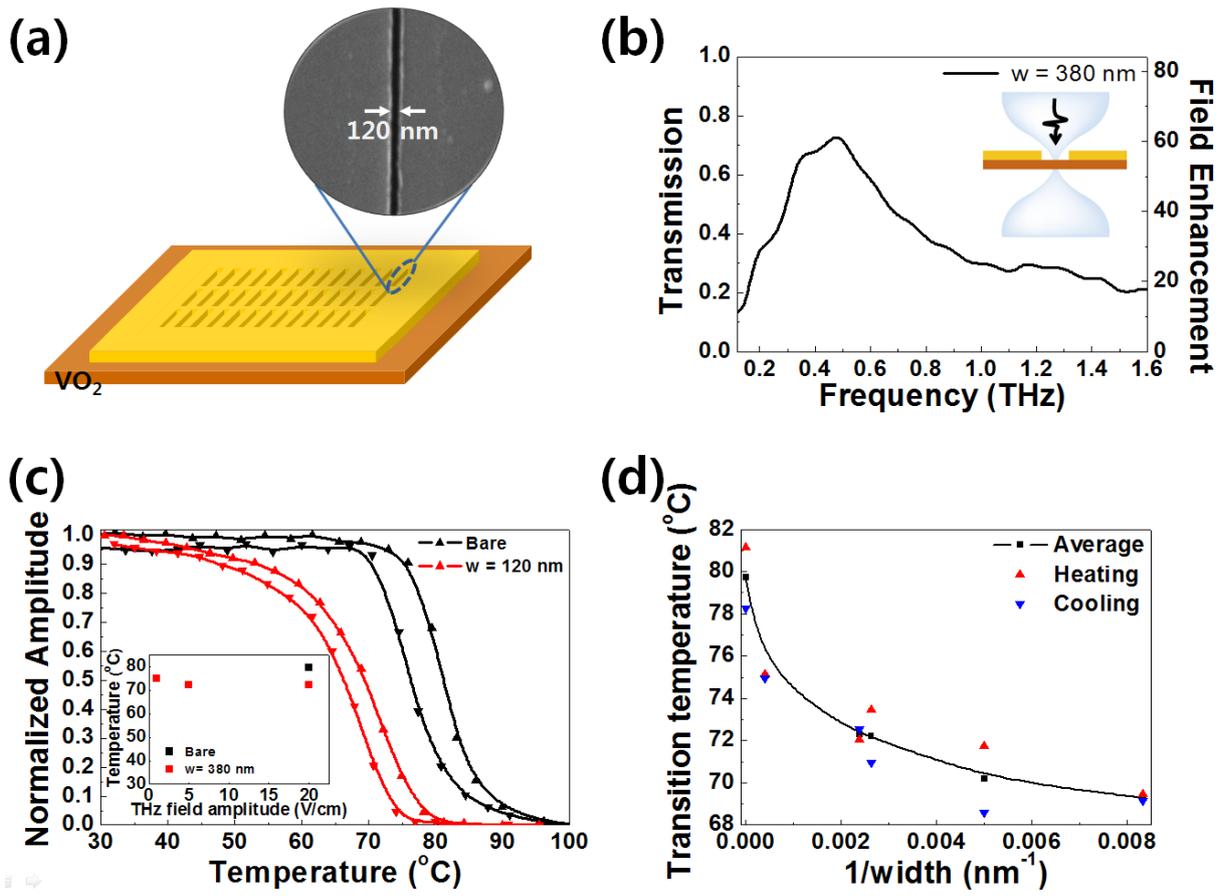

Fig. 2

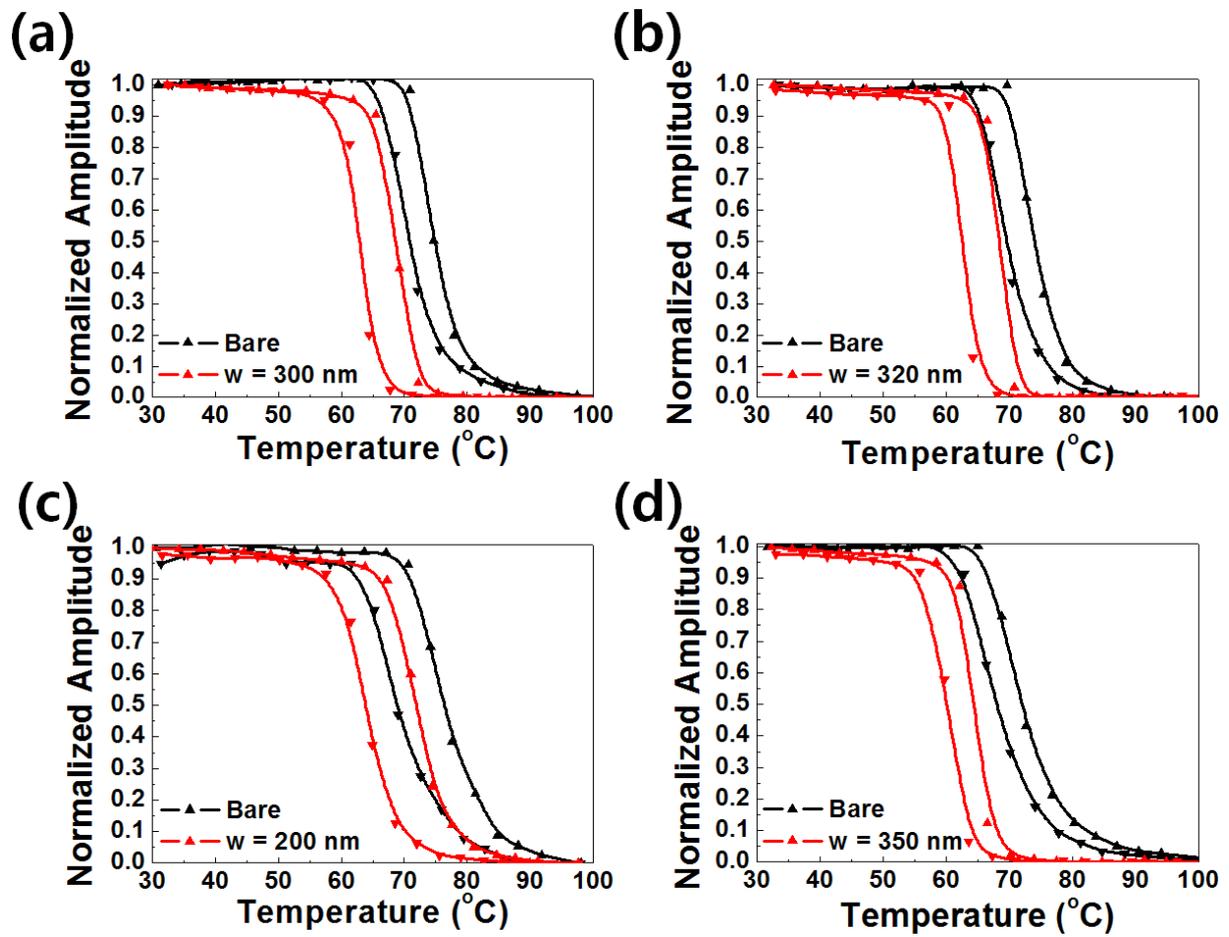

Fig. 3

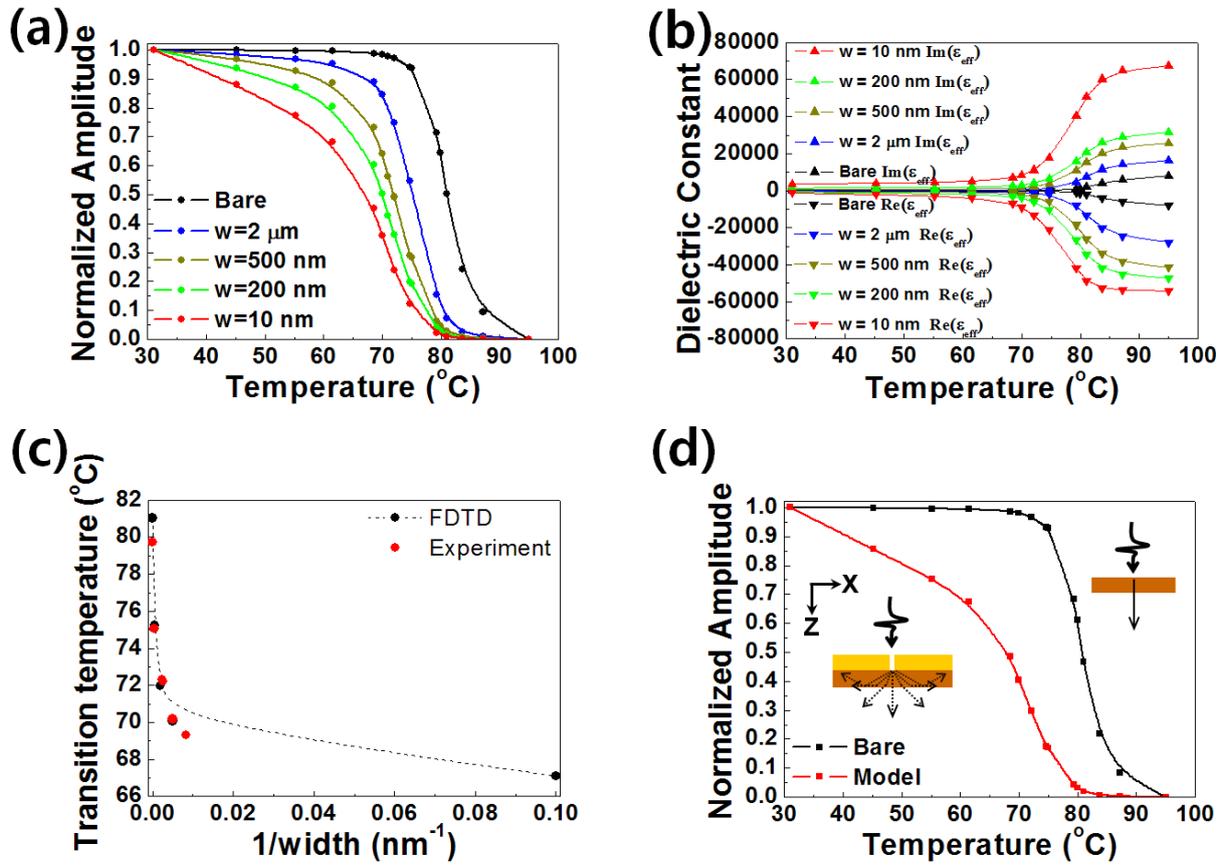